\documentclass[aps,prl,twocolumn,
superscriptaddress, 
nofootinbib]{revtex4-2}  
\usepackage{graphicx}  
\usepackage{dcolumn}   
\usepackage{bm,relsize}        
\usepackage{amssymb, amsmath,amsthm,mathtools}
\usepackage{slashed}   
\usepackage[colorlinks,citecolor=blue,urlcolor=violet,bookmarks=false,hypertexnames=true]{hyperref} 
\pdfoutput=1

\usepackage[usenames,dvipsnames,svgnames,table]{xcolor}

\def\beq{\begin{equation}}
\def\eeq{\end{equation}}

\def\MPl	{M_{\mathsmaller{\rm Pl}}}

\def\SM		{{\mathsmaller{\rm SM}}}
\def\FO		{{\mathsmaller{\rm FO}}}
\def\PT		{{\mathsmaller{\rm PT}}}
\def\OS		{{\mathsmaller{\rm OS}}}
\def\RH		{{\mathsmaller{\rm RH}}}
\def\D		{{\mathsmaller{\rm D}}}

\def\Csix   {C_6}

\begin{document}
\widetext
\leftline{DESY-24-007}

\title{\Large Early Matter Domination at Colliders: Long Live the Glueball!}
\author{Fady Bishara}
\affiliation{Deutsches Elektronen-Synchrotron DESY, Notkestr.~85, 22607, Hamburg, Germany}
\author{Filippo Sala}
\thanks{FS is on leave from LPTHE, CNRS \& Sorbonne Universit\'{e}, Paris, France.}
\affiliation{Dipartimento di Fisica e Astronomia, Università di Bologna
and INFN, Via Irnerio 46, I-40126 Bologna, Italy}
\author{Kai Schmidt-Hoberg}
\affiliation{Deutsches Elektronen-Synchrotron DESY, Notkestr. 85, 22607, Hamburg, Germany}

\date{\today}

\begin{abstract}
\noindent We prove that collider searches for long-lived particles (LLPs) can test
the dynamics responsible for matter domination in the early universe.
In this letter we concentrate on the specific example of glueballs from a GeV-scale confining dark sector and compute the dilution of cosmological relics induced by their decay. We then show that searches for long-lived glueballs from Higgs decays test increasing values of dilution at ATLAS and CMS, CODEX-b, ANUBIS and MATHUSLA. We identify the general features that make models of early matter domination discoverable via LLPs at colliders. Our study provides a quantitative physics motivation to test longer lifetimes. 
\end{abstract}

\maketitle

\paragraph{\bf Introduction.}
There is observational evidence that the energy budget of the universe was dominated by radiation when its age was roughly between one second and $10^5$ years, corresponding to temperatures of the thermal bath $\mathcal{O}$(MeV) $ > T > \mathcal{O}$(eV), and that matter started to dominate for $T < \mathcal{O}$(eV).
The form of energy that dominated at very high temperatures is unknown. 
It is then interesting to entertain the possibility that a phase of matter domination took place also at $T > \mathcal{O}$(MeV), and to look for ways to test it experimentally.
While the Standard Model (SM) in isolation predicts radiation domination at $T > \mathcal{O}$(MeV), a period of early matter domination (EMD) is instead 
predicted in some models that address shortcomings of the SM (see e.g.~\cite{Banks:1993en,deCarlos:1993wie,Moroi:1999zb}) and EMD has been invoked to dilute cosmological relics that would otherwise contradict observations, such as too heavy thermal dark matter~\cite{Giudice:2000ex}.

Concerning experimental tests of EMD,
to our knowledge, they are all indirect in the sense that they do not directly probe the dynamics that induces the EMD. Rather, they probe the  consequences of EMD on other obervables. 
Most of these tests come from the sky, for example via the EMD impact on the frequency spectrum of gravitational waves sourced before the EMD period (see e.g.~\cite{Gouttenoire:2019kij,Gouttenoire:2019rtn}) or on the energies of cosmic rays produced by dark matter annihilations~\cite{Cirelli:2018iax}.
And while there have been collider studies of scenarios which feature an EMD period, 
these do not test the dynamics at the origin of EMD, but rather its impact on independent physics, for example on dark matter freeze-in~\cite{Co:2015pka}.
It would be highly desirable to have more direct experimental tests of these dynamics. In this letter we show that EMD can be tested more directly in the laboratory via collider searches for the long-lived particles (LLPs) that induce it.

\medskip
\paragraph{\bf Dilution of relics.}
A particle that
\begin{itemize}
\item[i)] is not in equilibrium with the thermal bath while having a large number density (e.g. a particle that freezes out while relativistic)
\item[ii)] decays while non-relativistic
\end{itemize}
may end up dominating the energy budget of the universe and so realises EMD, until it decays and injects entropy into the SM bath.
This reduces the yield $n_{X}/s_\SM$ of any pre-existing relic $X$, where $n_{X}$ is its number density and $s_\SM$ the SM entropy, by a dilution factor (see e.g.~\cite{Cirelli:2018iax})
\begin{equation}
D_\SM
\simeq
\left[
1 + 0.77\, (g_\SM^\text{dec})^{1/3}
\Big(\frac{n^\FO}{s_\SM^\FO} \frac{m}{\sqrt{\Gamma M_\text{Pl}}} \Big)^\frac{4}{3}
\right]^\frac{3}{4},
\label{eq:DSM}
\end{equation}
where $M_\text{Pl} \simeq 2.4\times 10^{18}$~GeV is the reduced Planck mass, $\Gamma$ and $m$ are the width and mass of the decaying particle, $g^\text{dec}_\SM$ are the SM degrees of freedom at the time of decay, and $n^\FO/s_\SM^\FO$ is the yield of the decaying particle when it freezes out.
Eq.~(\ref{eq:DSM}) holds regardless of whether the dark and SM sectors have equal early temperatures or not.\footnote{
$D_\SM$ is the dilution of relics that are coupled to the SM, with respect to the case where the dark sector did not play a role. The dilution of relics coupled to both sectors, or to the dark one only, includes other factors of temperatures and degrees of freedom~\cite{Cirelli:2018iax}.
}

One may think that the long lifetimes tested in the lab (colliders, beam-dump experiments) are enough to directly test EMD. Point i) makes this less trivial: the SM-LLP couplings needed to produce enough LLPs in the lab are often large enough to keep them, in the early universe, in equilibrium with the bath until they become non-relativistic. Therefore their number density drops, so that they do not induce any dilution of relics.
Dark photons are a notable example of this case: a kinetic mixing that is small enough to give rise to a period of EMD implies negligible production at any collider, existing or foreseen (compare e.g. Fig.~7 of~\cite{Cirelli:2016rnw} with Fig.~10 of~\cite{Curtin:2014cca}). 
We next present a model that satisfies both i) and ii), induces a sizeable dilution, and is within reach of collider LLP searches.

\medskip

\paragraph{\bf Long-lived dark glueballs at colliders.}

We consider a dark confining gauge group $SU(N_\D)$ with no charged light dark fermions, such that the lightest composite states are glueballs. Without couplings to the SM the lightest of these glueballs is stable and a good DM candidate~\cite{Forestell:2016qhc,Acharya:2017szw}. Assuming the generation of small effective couplings to the visible sector, these glueballs will typically be long-lived and may lead to a phase of EMD as we show below.
The lowest dimensional portal with the SM is
\begin{equation}
    \mathcal{L}_6 =
    \Csix\, \alpha_\D \text{tr}G_{\mu\nu}G^{\mu\nu} \Phi^\dagger \Phi\,, \quad
    \Csix = \frac{y^2}{3\pi}\frac{1}{M^2}\,,
    \label{eq:Higgs_DarkGluons_dim6}
\end{equation}
where $\Phi$ is the SM Higgs doublet, $G_{\mu\nu}$ is the dark gauge field strength, $\alpha_\D = g^2_\D/4\pi$ with $g_\D$ the dark gauge coupling of $SU(N_\D)$.
For definiteness, we display the value of the Wilson coefficient $\Csix$ in the case where one has new vector-like fermions $F_{e,\ell}$ of mass $M$ in the fundamental of $SU(N_\D)$ and charged under the SM electroweak group $SU(2)\times U(1)$ such that the coupling $y \bar{F}_\ell F_e  \Phi$ is allowed, see~\cite{Juknevich:2009gg} for more details.
As defined, $\Csix$ does not run under $SU(N_\D)$.

The mass $m_0$ of the lightest glueball, which is a $0^{++}$ state, can be expressed in terms of the zero-flavour $\overline{\text{MS}}$ confinement scale $\Lambda_\D$ as\footnote{We have used $\Lambda_\D/\sqrt{\sigma} \simeq 0.5 + 0.3/N_\D^2$~\cite{Athenodorou:2021qvs} and our interpolation $m_0/\sqrt{\sigma} \simeq 3.07 (1+1/N_\D^2)$ of the lattice data in~\cite{Athenodorou:2021qvs}, where $\sigma$ is the string tension.
\label{foot:string} 
}
\beq
m_0
    \simeq \Big(6.1 + \frac{2.5}{N_\D^2}\Big) \Lambda_\D\,,
    \label{m0_Lambda}
\eeq
where here and in the following we neglect higher orders in $1/N_\D^2$.
To possibly have Higgs decays into glueballs one needs $m_h > 2 m_0$, where $h$ is the SM Higgs particle and $m_h \simeq 125.1$~GeV~\cite{ATLAS:2023oaq}. One then has $m_h \gg \Lambda_\D$, so that the Higgs decays into dark gluons that later hadronise into glueballs.
The Higgs width into dark gluons $A_\D$ reads
    \begin{equation}
        \Gamma(h \to A_\D A_\D) = (N^2_\D-1)\frac{\alpha^2_\D}{4\pi} \Csix^2 v^2 m_h^3\,,
        \label{eq:higgswidth_darkgluons}
    \end{equation}
    where $v\simeq246$~GeV and, in what follows, we use the two-loop expression~\cite{Larin:1993tp}
    $
   \alpha_\D(\mu)/4\pi
    = (\beta_0 \log\frac{\mu^2}{\Lambda_\D^2})^{-1} - \beta_1 \log\log\frac{\mu^2}{\Lambda_\D^2} (\beta_0^3 \log^2\frac{\mu^2}{\Lambda_\D^2})^{-1}
    $
    evaluated at $\mu=m_h/2$, with $\beta_0=11 N_\D/3$ and $\beta_1=34 N_\D^2/3$.

    Gluons then hadronise and produce glueballs. 
    We anticipate that, as in our parameter space $m_0$ will be sufficiently below the Higgs mass, this perturbative treatment is accurate and resonance effects are not relevant in the Higgs width of Eq.~\eqref{eq:higgswidth_darkgluons}.
    Dark shower simulations combined with a jet-like hadronisation model~\cite{Curtin:2022tou,Batz:2023zef} find that, for the glueball masses of interest for this study $4 \lesssim m_h/m_0 \lesssim 40$, the average number of glueballs resulting from $h \to A_\D A_\D$ is between 2 and 5.
    They also find that the most abundant glueball is the $0^{++}$ state $S_0$, immediately followed by the $2^{++}$ one, $S_2$, with mass $m_2 \simeq 1.5 m_0$~\cite{Athenodorou:2021qvs}.
    Other states are at least one order of magnitude rarer.

The portal in eq.~(\ref{eq:Higgs_DarkGluons_dim6}) induces $S_0$ decays via its mixing with the Higgs.
To compute the mixing we employ (see eqs (2.8) and (5.1) in~\cite{Juknevich:2009gg})
    \begin{equation}
        \langle 0| \alpha_\D \text{tr}G_{\mu\nu}G^{\mu\nu} | S_0 \rangle
        \equiv F^S_{0^{++}}
        = \frac{3.06}{4 \pi} m_0^3\,.
    \end{equation}
    At energies smaller than $\Lambda_\D$, the portal of eq.~(\ref{eq:Higgs_DarkGluons_dim6}) becomes
    \begin{equation}
        \mathcal{L}_{E\lesssim\Lambda_\D}^{6}
        = \frac{3.06}{4 \pi} \Csix \, m_0^3\, \Phi^\dagger \Phi S_0 
        \supset \frac{3.06}{4 \pi} \Csix \, m_0^3 \, v\, h S_0\,,
    \end{equation}
    thus inducing a Higgs mixing with $S_0$
    \begin{equation}
        \theta_{hS} = \frac{3.06}{4 \pi} \Csix \,\frac{m_0^3 \,v}{m_h^2-m_0^2}\,.
    \end{equation}
We then use 
\beq
\Gamma_0(m_0) = \theta_{hS}^2(m_0) \Gamma_h(m_0)\,,
\label{eq:glueball_width}
\eeq
where $\Gamma_h(m_0)$ is the width of a SM Higgs boson of mass $m_0$, which we take from~\cite{Winkler:2018qyg}.
    For $4~\text{GeV} < m_0 < m_h/2$, $\Gamma_0 = \Gamma_b + \Gamma_c +\Gamma_\tau$ to an excellent precision, where
    \begin{equation}
        \Gamma_{\!f} = N_f \frac{y_f^2\,\theta_{hS}^2}{16 \pi} m_0 \Big(1-\frac{4 (m_f^\OS)^2}{m_0^2}\Big)^{\!\frac{3}{2}}\,,
    \end{equation}
    $y_f = \sqrt{2}\,m_f/v$ ($m_f =$ 4.18, 1.78, 1.27~GeV respectively for $b,\tau, c$),
$m_f^\OS$ are the on-shell masses (= 4.78, 1.78, 1.67~GeV respectively for $b,\tau,c$)~\cite{Workman:2022ynf} $N_f = 3$ for quarks and $N_f = 1$ for charged leptons.
Coming to $S_2$, it decays to $S_0$ plus SM fermions via an off-shell Higgs boson, the associated width can be found in~\cite{Juknevich:2009gg} and is negligible with respect to $\Gamma_0$.
Dimension-8 operators, generated by the same vector-like fermions that generate Eq.~\eqref{eq:Higgs_DarkGluons_dim6}, open decay channels into gauge bosons~\cite{Juknevich:2009gg} which are negligible with respect to $\Gamma_0$.

\begin{figure*}[t]
\includegraphics[width=0.49\textwidth]{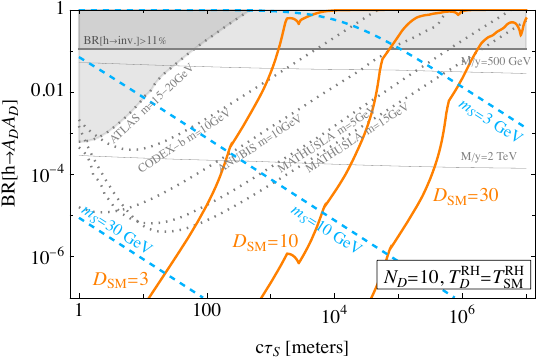}\;
\includegraphics[width=0.49\textwidth]{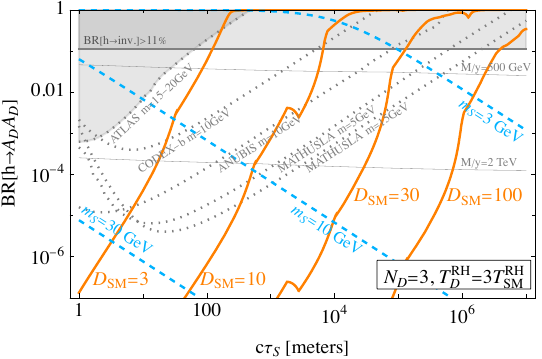}
\caption{\label{fig:HiggsDecays}
Higgs BR into dark gluons $A_\D$ from $SU(N_\D)$ versus lifetime of the lightest glueball.
We display contour lines of the mass of the lightest glueball (dashed blue), and of the dilution of early universe relics from glueball domination and decays (orange), $D_\SM$ of Eq.~\eqref{eq:DSM}, for $N_\D = 10 (3)$ and $T_\D^\RH/T_\SM^\RH = 1 (3)$ in the left-hand (right-hand) plot.
The horizontal shaded gray area is excluded at the LHC by limits on the Higgs invisible BR~\cite{ATLAS:2023tkt} (see also \cite{CMS:2023sdw}), searches for the new `quirky' EW fermions that induce the portal in Eq.~\eqref{eq:Higgs_DarkGluons_dim6} are weaker (for $y=1$). $M/y \sim f$ of fraternal Twin Higgs~\cite{Craig:2015pha}, so smaller BRs correspond to increasing tuning.
The dashed gray contours indicate limits or sensitivities from Higgs decays into LLPs, of fixed mass $m$, that result in displaced jets
i) at ATLAS~\cite{ATL-PHYS-PUB-2022-007} (CMS limits are comparable~\cite{CMS:2021juv}); 
ii) at Codex-b with 300 fb$^{-1}$~\cite{Gligorov:2017nwh,Aielli:2022awh} (searches in the CMS endcap muon detectors with 3~ab$^{-1}$ and the `higher $N_\text{hits}$ cut' have a comparable reach~\cite{Mitridate:2023tbj} for $c\tau \gtrsim 10$~m);
iii) ANUBIS~\cite{Bauer:2019vqk} with 3~ab$^{-1}$ using the `ceiling' 50-events line from~\cite{ANUBIS:2022} and
iv) MATHUSLA with 3~ab$^{-1}$ for $m=5$~\cite{Chou:2016lxi} and $15$~\cite{Curtin:2023skh} GeV (a boosted muon collider would have a better reach with a much smaller volume~\cite{Barducci:2023gdc}). 
See text for more details.
}
\end{figure*}

We now turn to experimental limits and sensitivities from displaced vertex searches on our scenario.
To summarise the previous discussion, in the parameter space of interest, decays of the SM Higgs into dark gluons result in a few long-lived dark glueballs, which mainly consist of $S_0$ and slightly less $S_2$. Dark glueballs then decay mostly to quark pairs.
In the parameter space where $S_0$ results in a displaced vertex at a given detector, $S_2$ results in missing energy.
We did not find interpretations in our scenario of searches for LLPs from Higgs decays across different (planned or existing) experiments\footnote{
The 36fb$^{-1}$ ATLAS search for one displaced vertex in the muon spectrometer plus missing energy (see the auxiliary material of~\cite{ATLAS:2018tup}) is the closest one we found to our expected signature. 
}, and recasting existing ones (as done e.g. in another LLP context in~\cite{CMS:2019qjk,Araz:2021akd}) goes well beyond the purpose of this paper.
For simplicity and consistency we then display in Fig.~\ref{fig:HiggsDecays} limits and sensitivities from searches for Higgs decays to two LLPs, which in turn result in displaced jets.
Experiments present the results of these searches for fixed LLP masses, while our LLP mass $m_0$ varies as a function of $c\tau$ and of the Higgs BR.
Their reach will turn out to cut into the parameter space of sizeable dilutions by orders of magnitude in $c\tau$ and BR[h$\to A_\D A_\D$], so that our qualitative conclusions will be solid even though the shown sensitivities should only be considered indicative.

In Fig.~\ref{fig:HiggsDecays} we also display contours of glueball mass $m_0$ obtained by substituting $y/M$ for $\Gamma_0$ via Eq.~(\ref{eq:higgswidth_darkgluons}). Eq.~\eqref{eq:Higgs_DarkGluons_dim6} induces the negligible mass correction $\Delta m_0^2 \sim 10^{-3} m_0^2 (y \text{TeV}/M)^2$, where we have used $\langle S_0| 4\pi \alpha_\D \text{tr}G_{\mu\nu}G^{\mu\nu} | S_0 \rangle \sim m_0^2$. 

In Fig.~\ref{fig:HiggsDecays} we also display contours of $M/y$. This is useful in two respects.
Firstly, $M/y \sim f$, where $f$ is the symmetry-breaking scale of the fraternal Twin Higgs~\cite{Craig:2015pha}, which is one of the UV motivations of our scenario. There the tuning of the weak scale is proportional to $f^2/v^2$, so that smaller values of BR[h$\to A_\D A_\D$] correspond to a larger fine-tuning.  
Secondly, knowing the values of $M/y$ allows to gauge the impact of collider searches for the heavy vector-like fermions of our chosen UV completion of Eq.~(\ref{eq:Higgs_DarkGluons_dim6}). As the strings of the dark confining group do not have any light dark quark to break into, these heavy fermions behave as `quirks'~\cite{Kang:2008ea}, so that searches for vector-like leptons as performed in~\cite{CMS:2022cpe} do not apply. For the GeV values of the confining scale we are interested in here, we are not aware of collider searches targeting them.
Following~\cite{Knapen:2017kly,Evans:2018jmd}, one could estimate the LHC reach for quirks with the one of monojet searches, as roughly 200~GeV. This would however be too aggressive, as for GeV confining scales quirks annihilate on detector scales and so evade the typical cuts of monojet searches. Since anyway, for $y=1$, this aggressive reach is weaker than limits from invisible Higgs decays into dark glueballs~\cite{ATLAS:2023tkt,CMS:2023sdw}, we only display the latter in Fig.~\ref{fig:HiggsDecays}.

\medskip

\paragraph{\bf Cosmology down to the phase transition.}
 Having discussed the parameters of the model and the collider phenomenology of our interest, we now switch to its cosmology, with the final aim of computing the dilution of relics induced by glueball decays in the early universe via Eq.~(\ref{eq:DSM}). In particular, the computation of $n^\FO/s_\SM^\FO$ for glueballs requires the knowledge of the evolution of the dark sector in the early universe, including whether or not it equilibrates with the SM bath and at which temperatures: as anticipated in the introduction, equilibrium with the SM bath would reduce the number density of glueballs and hence the dilution they induce upon decay.
 
 As the universe cools,
$SU(N_\D)$ will experience a confining phase transition (PT) at a temperature\footnote{
We use $T^c_\D
\simeq (0.59+0.46/N_\D^2)\sqrt{\sigma}$~\cite{Lucini:2012wq}, footnote~\ref{foot:string} and Eq.~(\ref{m0_Lambda}).
}
\beq
T^\PT_\D
\simeq T^c_\D
\simeq \Big(1.2 + \frac{0.21}{N_\D^2}\Big) \Lambda_\D\,
\simeq \frac{m_0}{5.1+1.2/N_\D^2},
\label{eq:Tc}
\eeq
where $T^c_\D$ is the critical temperature at which deconfinement happens and $T^\PT_\D
\simeq T^c_\D$ is an assumption justified as long as the PT is not significantly supercooled, see e.g.~\cite{Gouttenoire:2022gwi}.
For temperatures larger than a few GeV, the fastest process that could maintain chemical and kinetic equilibrium involves bottom and charm quarks due to the Yukawa-like coupling structure.
The relevant cross section for chemical equilibrium, averaged over initial degrees of freedom, reads  ($s$ is the usual Mandelstam variable)
\begin{equation}
\begin{split}
    \sigma v (A_\D A_\D \rightarrow q\bar{q})
    &\simeq \frac{3}{N_\D^2} \frac{\alpha_\D^2\,C_6^2\, y_q^2}{4\pi} \frac{s^2}{m_h^4} v^2 \\
    \end{split}
    \label{eq:sigmav_largeT}\,.
\end{equation}
This reaction is the relevant one to consider since it suppresses the $A_\D$ abundance.
We find that the dark sector chemically decouples from the SM, $n_\D \sigma v/H < 1$, if the dark temperature satisfies the inequality
\begin{equation}
    T_{\D} < T_\text{eq}^\text{rel} \equiv 12~\text{GeV}
    \Big(\frac{M/y}{\rm TeV}\Big)^{\!\frac{4}{5}} 
    \Big(\frac{0.5}{\alpha_\D N_\D}\Big)^{\!\frac{2}{5}} 
    \Big(\frac{N_\D}{3}\Big)^{\!\frac{3}{5}}\,,
    \label{eq:NOeq_abovePT}
\end{equation}
where we have used $n_\D = \zeta(3) g_\D T_\D^3/\pi^2$, $s \sim T_\D^2$, $g_\D \simeq 2 N_{\D}^2$ and where $H$ denotes the Hubble parameter.
For simplicity we use $ H \simeq \pi  \sqrt{g_\D/90} \,T_\D^2/M_\text{Pl}$, ignoring the contribution of the SM bath which, for the values of $N_\D$ that we will choose and for the temperatures of Eq.~(\ref{eq:NOeq_abovePT}), is at most comparable to the one of the dark sector.
This approximation is conservative in our context, because it results in a slightly lower value of $T_\text{eq}^\text{rel}$ than the one obtained including also the SM bath, and it will then correspond slightly smaller dilutions.
Note that Eq.~(\ref{eq:NOeq_abovePT}) defines $T_\text{eq}^\text{rel}$, that $\alpha_\D N_\D$ is roughly constant in $N_\D$ and $\alpha_\D N_\D \lesssim 0.5$ for $m_0 \lesssim 30$~GeV, and that $T_\text{eq}^\text{rel} \propto (\sigma v)^{\!\frac{1}{5}}$. The temperature of kinetic equilibration is at most equal to $T_\text{eq}^\text{rel}$ up to $O(1)$ factors.

In what follows we will consider two cases:
\begin{itemize}
    \item[i)]
    the dark and SM sectors are in equilibrium at high temperature, $T_\D = T_\SM$, so that Eq.~(\ref{eq:NOeq_abovePT}) indicates the common temperature when they decouple;
    \item[ii)]
    inflation reheats the SM at a lower temperature than the dark sector, $T_\SM^\RH < T_\D^\RH$, so that $T_\text{eq}^\text{rel}$ of Eq.~(\ref{eq:NOeq_abovePT}) indicates the largest $T_\D^\RH$ for which ii) holds (of course we need also $T_\D^\RH > T_\PT$).
\end{itemize}
Regardless of the reheating assumptions, Eqs~\eqref{eq:NOeq_abovePT} and~\eqref{eq:Tc} inform us $T_\D^\PT < T_\text{eq}^\text{rel}$ for $m_0 < 30$~GeV, where we have used that the limit on the Higgs invisible BR~\cite{ATLAS:2023tkt} corresponds roughly to $M/y > 400$~GeV. So the SM and the dark sector will always be decoupled at the PT in the parameter space of our interest. 
The abundance of vector-like quarks, that generate the coupling in Eq.~(\ref{eq:Higgs_DarkGluons_dim6}), is depleted by the time of the PT by their annihilations into the $A_\D$. They could have an impact on glueball cosmology if they were much heavier, see e.g.~\cite{Soni:2017nlm}.

At the confining PT the dark gluons are converted into a large number of glueballs.
Their number density $n_0^\PT$, just after the PT but before they start a `cannibalistic' period where $3\to2$ interactions are active, can be simply estimated by energy conservation, as also done in~\cite{Redi:2020ffc},
as
\beq
m_0 n_0^\PT = \rho^\D_\PT + L_h\,,
\eeq
where $\rho^\D_\PT  = \pi^2 g_\D T_\PT^4/30$ is the energy density of the dark sector at the PT temperature $T_\PT$, and $L_h$ is the latent heat of the PT.\footnote{We actually expect the dark energy (and entropy) densities to depart from their equilibrium values soon before the PT, because the degrees of freedom start to rearrange themselves due to the strong coupling. This expectation is possibly backed by lattice computations like~\cite{Borsanyi:2012ve}.
The fact that the SM and dark sectors are decoupled at the PT anyway supports quantitatively our assumption, that this decrease in dark degrees of freedom does not result in a transfer of energy or entropy into the SM bath.
}
To make contact with an approximation we will employ later on, we have also assumed for simplicity that all glueballs have mass $m_0$.
Using~\cite{Datta:2010sq} 
\beq
L_h \simeq (N_\D^2-1) \,T_c^4\,\Big(0.39-\frac{1.6}{N_\D^2}\Big)\,,
\eeq
with $T_c$ of Eq.~(\ref{eq:Tc}), we find
\begin{equation}
\label{eq:nSPT}
    n_0^\PT \simeq
    \frac{\rho_\D^\PT + L_h}{m_0}
    \simeq 10^{-3}\,\Big(1.5 \,N_\D^2 - 5.1 + \frac{7.4}{N_\D^2}\Big) m_0^3
\end{equation}
corresponding e.g. to $n_0^\PT = 9.4 \cdot 10^{-3}m_0^3~(0.15 \,m_0^3)$ for $N_\D = 3~(10)$.
Note that $n_0^\PT \gg n_0^{\rm eq} \sim (m_0 T_\PT/(2\pi))^{\!\frac{3}{2}} e^{-m_0/T_\PT}$. In addition to energy conservation, this can be understood qualitatively with the fact that the number of degrees of freedom of excited glueballs $S^*$, the `Hagedorn spectrum', scales as $e^{m_{S^*}/\Lambda}$ (see~\cite{Meyer:2009tq,Borsanyi:2012ve} for quantitative uses of this relation in the successful determination of the entropy density at the PT).
Assuming entropy rather than energy conservation, we obtain marginally different values for the number density $n_0^\PT$.

\medskip
\paragraph{\bf Cosmology of long-lived dark glueballs.}
In order to estimate the number density of glueballs $n^\FO_S$, that enters Eq.~(\ref{eq:DSM}) for the dilution factor, we need to account for their evolution until their number-changing interactions freeze out.
The fastest ones are $2 \to 2$, which change the relative abundances of different glueball states, followed by $3 \to 2$ `cannibalistic' interactions, which change the overall number of glueballs.
Here we adopt a simplified single-state model consisting only of~$S_0$. Ref.~\cite{Forestell:2016qhc} has proven that this reproduces very well the $S_0$ abundance obtained from a more complete treatment including all glueballs.
Our simplified model then describes well also the EMD phase, because $2 \to 2$ interactions make the abundance of heavier glueballs extremely Boltzmann suppressed~\cite{Forestell:2016qhc}, even the one of those like $0^{-+}$ and $1^{+-}$ that would be stable if only the portal of Eq.~\eqref{eq:Higgs_DarkGluons_dim6} existed. 
In order to compute $n^\FO_S$, we use that the dark sector and the SM are not in equilibrium neither before the PT (valid for small enough $T_\PT$, see Eq.~(\ref{eq:NOeq_abovePT})) nor after the PT (we will prove this later). Therefore their comoving entropy densities are conserved, and so is the ratio of their entropy densities
\begin{equation}
\label{eq:R}
R \equiv \frac{s_\D}{s_\SM}
=\frac{s_\D}{s_\SM}\Big|_\PT
= 
\frac{2 (N_\D^2-1)}{g_\SM^\PT} \Big(\frac{T_\D^\PT}{T_\SM^\PT}\Big)^3\,.
\end{equation}
After the PT number-changing glueball interactions are very fast, so their chemical potential quickly relaxes to zero and $s_
\D = m_0 n_\D/T_\D$. One then obtains
\begin{equation}
\label{eq:yield_glueballs}
\frac{n_0}{s_\SM} = R \frac{T_\D}{m_0}\,,
\end{equation}
which in particular holds at freeze-out (not later, because then chemical potentials will be different from zero), so that $n_0/s_\SM$ corresponds to $n^\FO/s_\SM^\FO$ of Eq~(\ref{eq:DSM}).

One is left with the computation of $x^\FO_\D \equiv m_0/T_\D^\FO$. We calculate it from the instantaneous approximation relation (see e.g.~\cite{Carlson:1992fn})
\begin{equation}
3 H_\FO = \langle \sigma_{32} v^2 \rangle (n_0^\FO)^2 x^\FO_\D\,,
\end{equation}
where $\langle \sigma_{32} v^2 \rangle \simeq \frac{1}{(4\pi)^3} (\frac{4\pi}{N_\D})^6 \frac{1}{m_0^5}$ (see e.g.~\cite{Forestell:2016qhc}, and~\cite{Yamanaka:2019yek} for a lattice calculation of $\sigma_{22}$ supporting our $\langle \sigma_{32} v^2 \rangle$ estimate)
and $n_0^\FO = (m_0 T_\D/(2\pi))^\frac{3}{2} \exp(-m_0/T_\D)$.
If the universe's energy density at glueball freeze-out is dominated by the glueballs, then $H_\FO\simeq (m_0 n_0^\FO/3)^\frac{1}{2}/\MPl$ and we obtain
\begin{equation}
\label{eq:xFOD}
 e^{2 x_\D^\FO} (x_\D^\FO)^\frac{5}{3}   
 = X
 \equiv
 \frac{32\pi}{3^\frac{2}{3}} \frac{1}{N_\D^8} \Big(\frac{\MPl}{m_0}\Big)^\frac{4}{3}\,.
\end{equation}
For the values of $N_\D$ and $m_0$ of our interest, we find $16 \lesssim x_\D^\FO \lesssim 22$.
In the limit $X \gg 1$ the solution of Eq.~(\ref{eq:xFOD}) is approximated by $x_\D^\FO = \frac{1}{2}\log X$ to better than 20\%, still we use the full solution for our numerical results and figures.
The ratio between the dark and SM energy densities at FO, in case i) where the sectors were at equilibrium at early temperatures, is $\rho_\D^\FO/\rho_\SM^\FO \simeq 10^2 (g_\SM^\FO)^\frac{1}{3} (\frac{N_\D^2-1}{g_\SM^\PT})^\frac{4}{3}$ for $x_\D^\FO = 15$, and it is much larger for larger $x_\D^\FO$. This confirms the validity of our assumption in the derivation of $x_\D^\FO$ via Eq.~(\ref{eq:xFOD}), both in case i) and of course in case ii), where the energy density of the SM is even smaller.

We finally plug the solution $x_\D^\FO = m_0/T_\D^\FO$ of Eq.~(\ref{eq:xFOD}) into Eq.~(\ref{eq:yield_glueballs}) for $n_0/s_\SM$, with $R$ from Eq.~(\ref{eq:R}). In case i) we work with $T_\D^\PT = T_\SM^\PT$ and $g_\SM^\PT = 75$ and in case ii) with $T_\D^\PT = 3\,T_\SM^\PT$ and $g_\SM^\PT = 50$, where we keep a fixed $g_\SM^\PT$ for simplicity, including variations has a minor impact on our dilution lines.
We then obtain the contour lines of dilution factor $D_\SM$ of Eq.~(\ref{eq:DSM}) shown in Fig.~\ref{fig:HiggsDecays}.
We have found that $n_0/s_\SM$ (and therefore $D_\SM$) is suppressed, with respect to the one in absence of cannibalism, by a log factor. 
Our result can be qualitatively understood by noting that the energy density of a cannibal bath decreases faster than that of a frozen-out bath, because cannibalism converts mass into temperature, which redshifts, while mass doesn't.
Therefore one expects decays of particles that underwent cannibalism to induce less dilution than decays of particles that didn't\footnote{We thank Fatih Ertas, Felix Kahlhoefer and Carlo Tasillo for offering us this insightful explanation.}, as quantified for the first time in~\cite{Ertas:2021xeh} in another model.\footnote{
Contrary to our picture, in~\cite{Ertas:2021xeh} cannibalism ends before the cannibals dominate the universe's energy.
}
The glueballs then decay and inject entropy in the SM bath, diluting pre-existing relics. For the decay lengths that we consider, they do so at $T_\SM >$ few MeV and so do not spoil big bang nucleosynthesis~\cite{Hasegawa:2019jsa}.

\medskip

Finally, we check that the dark sector does not go back to equilibrium with the SM after the PT, because that was our hypothesis in deriving Eq.~(\ref{eq:yield_glueballs}), and because if that happened then the glueballs number density would become exponentially suppressed before they decay, and so not realise any dilution.
The cross section of the dominant number-changing process reads
 \begin{equation}
     \sigma v(S_0 S_0 \to q\bar{q})
     = \frac{3}{2} \frac{\,y_q^2 \, \theta_{hS}^2 \,\lambda_3^2}{4\pi\,m_0^4} \Big(1-\Big[\frac{m_q^\OS}{m_0}\Big]^2\Big)^{\!\frac{3}{2}},
     \label{eq:sigmav_glueglue}
 \end{equation}
 where $\lambda_3 \sim 4\pi \Lambda_\D/N_\D$ is the effective coupling between three $S_0$ and $q$ refers to the heaviest quarks that the glueballs can annihilate into.
 We find that $n_0 \sigma v < H$ after the PT is satisfied if
 \begin{equation}
     m_0 \lesssim 13~\text{GeV}
     \Big(\frac{M/y}{\rm TeV}\Big)^{\!\frac{4}{5}}
     \Big(\frac{y_q}{y_b}\Big)^{\!\frac{2}{5}}
     \Big(\frac{N_\D}{3}\Big)^{\!\frac{1}{5}}\,,
     \label{eq:mS_afterPT}
 \end{equation}
where for simplicity and to be conservative we have used $H = H_\D \simeq \sqrt{m_0 n_0/3}/M_\text{Pl}$ and $n_0^\PT$ of Eq.~(\ref{eq:nSPT}), and omitted the factor
$(1-(m_q^\OS)^2/m_0^2)^{\!\frac{3}{2}}$
in Eq.~(\ref{eq:sigmav_glueglue}).
We find that the condition of non-equilibration after the PT, Eq.~(\ref{eq:mS_afterPT}), is weaker than either the LHC limit on the Higgs invisible BR~\cite{ATLAS:2023tkt} or the ATLAS LLP limits~\cite{ATLAS:2018tup}, so we do not display it in Fig.~\ref{fig:HiggsDecays} to avoid clutter. Our calculation of $D_\SM$ is then valid in the white area.

\medskip

\paragraph{{\bf Discussion and Outlook.}}
In this letter we proved that long-lived particle (LLP) searches at colliders can directly test 
the dynamics responsible for the dilution of early-universe relics via a period of early matter domination (EMD).
These models were so far testable only indirectly via their imprints on cosmological relics or cosmic rays: our study thus opens a new avenue to test them directly in the laboratory. Conversely, it provides a novel quantitative physics case for LLP searches, by mapping their reach to dilution factors of relics in the early universe.

The generic non-trivial requirement, to be able to test the EMD dynamics with LLPs at colliders, is to simultaneously achieve:
\begin{itemize}
    \item[a)] the long lifetimes needed both for  EMD and for LLP collider searches;
    \item[b)] sizeable LLP collider production;
    \item[c)] a large cosmological abundance of out-of-equilibrium LLPs at the time of their decay.
\end{itemize}
In this paper we achieved this in a $SU(N_\D)$ confining group with no light fermions where the LLPs are the dark glueballs, whose abundance is set by the confining phase transition.
%
Another possibility, to satisfy a), b), and c), which we leave for future work, is producing the LLP at colliders from the decay of one or more new particles that do not sizeably couple the LLP to the cosmological bath. 

Our results motivate further studies of glueball cosmology and of their searches at colliders, and in general collider searches for not-too-forward LLPs.

\medskip

\subsection{Acknowledgements}

We thank Szabolcs Borsanyi, Frederik Depta, Fatih Ertas, Ed Hardy, Felix Kahlhoefer, Felix Kling, Thomas Konstandin, Giacomo Landini, Juri Smirnov, Pedro Schwaller, Carlo Tasillo and Andrea Tesi for useful discussions, and David Curtin and Michele Redi for feedback on the manuscript.
\medskip

\footnotesize
\noindent Funding and research infrastructure acknowledgements: 
\begin{itemize}
\item[$\ast$] F.S. acknowledges support from the Italian INFN program on Theoretical Astroparticle Physics (TAsP) and from the French Agence Nationale de la Recherche (ANR) under grant ANR-21-CE31-0013 project DMwithLLPatLHC.
F.S. thanks INFN Florence for hospitality while this work was being completed.
\item[$\ast$]
K.S.\ and F.B.\ acknowledge support from the Deutsche Forschungsgemeinschaft (DFG, German Research Foundation) under Germany’s Excellence Strategy – EXC 2121 “Quantum Universe” – 390833306.
\item[$\ast$] We all also acknowledge support
by COST (European Cooperation in Science and Technology) via the COST Action COSMIC WISPers CA21106.
\end{itemize}

\bibliography{EarlyMatterDom_Lab}
\bibliographystyle{JHEP}
\end{document}